\documentclass[a4paper]{article}

\usepackage{INTERSPEECH2021}
\usepackage{bm}
\usepackage{graphicx}
\usepackage{float}
\usepackage{ascmac} 
\usepackage{fancybox}
\usepackage{float}
\usepackage{tabularx}
\usepackage{booktabs}
\usepackage{amssymb}
\usepackage{multirow}
\usepackage{cite}
\usepackage{multirow}
\usepackage[table,xcdraw]{xcolor}
\usepackage[normalem]{ulem}
\useunder{\uline}{\ul}{}
\usepackage{url}

\AtBeginDocument{
  \abovedisplayskip     =.8\abovedisplayskip
  \abovedisplayshortskip=.8\abovedisplayshortskip
  \belowdisplayskip     =.8\belowdisplayskip
  \belowdisplayshortskip=.8\belowdisplayshortskip}

\title{Should We Always Separate?: Switching Between Enhanced and Observed Signals for Overlapping Speech Recognition}
\name{Hiroshi Sato, Tsubasa Ochiai, Marc Delcroix, \\Keisuke Kinoshita, Takafumi Moriya, Naoyuki Kamo}

\address{NTT Corporation, Japan}
\email{hiroshi.satou.bh@hco.ntt.co.jp}

\ninept

\begin{document}

\maketitle
\begin{abstract} 
Although recent advances in deep learning technology improved automatic speech recognition (ASR), it remains difficult to recognize speech when it overlaps other people's voices. 
Speech separation or extraction is often used as a front-end to ASR to handle such overlapping speech.
However, deep neural network-based speech enhancement can generate `processing artifacts' as a side effect of the enhancement, which degrades ASR performance.
For example, it is well known that single-channel noise reduction for non-speech noise (non-overlapping speech) often does not improve ASR.
Likewise, the processing artifacts may also be detrimental to ASR in some conditions when processing overlapping speech with a separation/extraction method, although it is usually believed that separation/extraction improves ASR. 
In order to answer the question `Do we always have to separate/extract speech from mixtures?', we analyze ASR performance on observed and enhanced speech at various noise and interference conditions, and show that speech enhancement degrades ASR under some conditions even for overlapping speech.
Based on these findings, we propose a simple switching algorithm between observed and enhanced speech based on the estimated signal-to-interference ratio and signal-to-noise ratio.
We demonstrated experimentally that such a simple switching mechanism can improve recognition performance when processing artifacts are detrimental to ASR.

\end{abstract}

\section{Introduction}
Recent advances in deep learning technology greatly improved the ability of speech recognition to recognize speech \cite{hinton2012deep,saon2017english}. 
However, it remains challenging to recognize speech under severe conditions such as speech overlapping with interfering speech (speech noise) \cite{barker2018fifth} and/or background noise (non-speech noise) \cite{barker2015third}.
To mitigate the adverse effect of interfering speech and noise, speech-enhancement technologies, including speech separation, target speech extraction, and noise reduction have widely been introduced and investigated as a front-end of ASR~\cite{qian2018single,delcroix2018single,li2014overview}.

\emph{Speech separation} and \emph{speech extraction} are proposed to deal with an overlapping speech of multiple speakers.
Speech separation~\cite{comon2010handbook,wang2018supervised} enables speech recognition of each speaker's voice by separating all speakers in the observed speech mixture.
Target speech extraction~\cite{vzmolikova2019speakerbeam} enables the recognition of the target speaker's voice by extracting only his/her voice from the observed mixture.
Various approaches have been proposed for speech separation and extraction so far, and even in single-channel setup, they realized drastic improvement on overlapping speech recognition by dealing with interfering speech~\cite{qian2018single,delcroix2018single,kanda2019auxiliary,denisov2019end,delcroix2019end}.

\emph{Noise reduction} is another technology to deal mainly with background noise.
For multichannel setup, beamformer has been shown to be very effective to improve ASR performance because it introduces only few `processing artifacts'~\cite{barker2015third,heymann2016neural,sainath2017multichannel}.
On the other hand, for single-channel setup, it is well-known that noise reduction technologies tend to degrade ASR performance due to the processing artifacts introduced by non-linear transformations in neural network\cite{yoshioka2015ntt,chen2018building,fujimoto2019one}.
Especially recent ASR systems trained on data covering various noise conditions tend to be less affected by non-speech noise than by the artifacts.

Although many have reported performance improvement using speech separation or extraction in overlapping conditions, we wonder if in some conditions the artifacts induced by these enhancement methods may be detrimental to ASR. 
In other words, should we always separate/extract speech from mixtures? 
In this work, we focus on the single-channel speech enhancement, especially single-channel `speech extraction' problem.
To investigate the above question, we evaluate the ASR performance on the `observed mixture' and `enhanced speech' at various noise and interference conditions. 
As a result, we show that depending on the balance of signal-to-interference ratio (SIR) and signal-to-noise ratio (SNR), ASR performance on observed mixtures outperforms that on enhanced speech.
Besides, we also demonstrate that we can improve ASR performance by simply switching between the observed mixture and enhanced speech based on estimated SIR and SNR.

Indeed speech separation/extraction is a strong way to deal with overlapping speech, but the results suggest that we should consider where to apply it.
The results obtained in this work showed a new strategy for further improving ASR performance for overlapping speech and present a new direction of speech separation/extraction research as the ASR front-end.

\section{Related Works}
In this work, we propose a switching method between observed mixture and enhanced speech for overlapping speech.
Similarly, a preceding work called Voice Filter Light\cite{wangvoicefilter} switched observed mixture and enhanced speech to improve ASR results.

In the preceding work, they improved the ASR performance on the non-overlapping region by decreasing the enhancement strength for the absence of interfering speakers, according to the assumption that speech extraction hurt ASR performance for non-overlapping regions.

Indeed both preceding work and our work share the common idea of switching between enhanced and observed speech, they are different in that the preceding work used the observed mixture for \emph{non-overlapping region}, while our work use it even for \emph{overlapping region}.
Therefore, both approaches would be complementary.

Although we conducted experiments on the almost fully overlapping dataset to validate the concept, we plan to investigate the application for the partially overlapping dataset by combining our proposal with switching for non-overlapping region\cite{wangvoicefilter} in future works.

Other approaches to mitigate the effect of processing artifacts include re-training of the ASR module on enhanced speech or performing joint-training of the speech enhancement and ASR.
Unlike our proposed method, these approaches often imply modifying the ASR back-end, which may not be possible in many practical applications.
Besides, to the best of our knowledge, there has not been extensive investigation comparing the performance of the enhancement with the observed mixture for various conditions for joint-training of speech separation/extraction and ASR, and thus the proposed switching approach may still be useful in some conditions.
Investigation of the switching mechanism in joint-training conditions including optimization of the switching parameters with ASR criterion will be part of our future works.

\section{Proposed Method}
In this section, we introduce the ASR pipeline that we use for our experiments and describe the proposed switching approach. 

\subsection{Overview of target speech recognition pipeline}
In this paper, we assume that a speech signal is captured by a single microphone. 
The single-channel observed mixture $\mathbf{Y} \in \mathcal{R}^{T}$ is given by:
\begin{align}
    \mathbf{Y} = \mathbf{S} + \mathbf{I} + \mathbf{N},
\end{align}
where $\mathbf{S}$ denotes the target signal, $\mathbf{I}$ denotes a signal that contains interfering speakers, and $\mathbf{N}$ denotes the background noise. $T$ is the number of samples of the observed waveform.

The target speech extraction framework aims at extracting the signal $\mathbf{S}$ of the target speaker from the observed mixture $\mathbf{Y}$ \cite{qian2018single,delcroix2018single}.
In target speech extraction, it is assumed that an `auxiliary information' about the target speaker is available to inform the network of which speaker to extract. 
Hereafter the auxiliary information is referred to as `target speaker clues'.
The target speaker clue is typically given as a waveform of a pre-recorded enrollment utterance of the target speaker, which is denoted as $\mathbf{C}^{\text{S}} \in \mathcal{R}^\tau$ where $\tau$ is the number of samples of the enrollment waveform.
Given the observed mixture and the target speaker clue, the target signal is estimated as follows:
\begin{align}
    \label{eq:spkbeam}
    \hat{\mathbf{S}} = \text{SE}(\mathbf{Y}, \mathbf{C}^{\text{S}})
\end{align}
where $\hat{\mathbf{S}}$ denotes the estimated signal of the target speaker 
and $\text{SE}(\cdot)$ denotes the speech extraction module. 

The target speech extraction can be utilized as a front-end for ASR to recognize the target speech in the mixture.
In ASR pipeline for the target speaker, the extracted signal $\hat{\mathbf{S}}$ is used as an input to the ASR module as follows:
\begin{align}
    \hat{\mathbf{W}} = \text{ASR}(\hat{\mathbf{S}}),
\end{align}
where $\hat{\mathbf{W}}$ denotes the estimated transcription of the target speaker and $\text{ASR}(\cdot)$ denotes the ASR module, which transforms the waveform into its corresponding transcription.

\begin{figure}[tb]
 \begin{center}
  \includegraphics[width=\hsize]{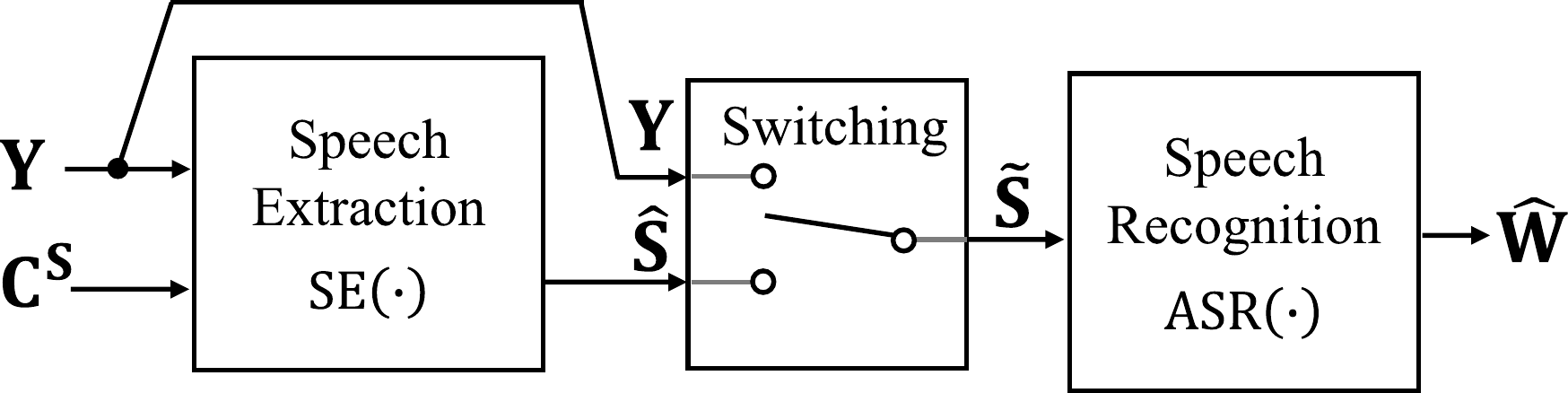}
 \end{center}
 \vspace{-12pt}
 \caption{The overview of ASR pipeline with proposed switching.}
 \vspace{-12pt}
 \label{fig:framework}
\end{figure}

\subsection{SNR and SIR-based input signal switching}
Target speech extraction benefits ASR by removing interfering speech that hinders the ASR performance even for ASR models trained robust for noise.
However, it is possible that the target speech extraction is detrimental to ASR in some conditions, because of 1) processing artifacts produced by non-linear transformations in target speech extraction and 2) the capability of ASR to recognize the dominant speaker.
It is empirically known that ASR can recognize only the dominant speaker even under the presence of the interfering speaker under relatively moderate SIR conditions.

In our experiment, we actually observed that sometimes ASR can perform well on observed mixture $\mathbf{Y}$ than on enhanced speech $\hat{\mathbf{S}}$ according to the balance between SIR and SNR (see experiments in Section~\ref{section:analysis}).
This finding suggests that we can improve ASR performance if we can correctly switch between the observed mixture and enhanced speech for the input of ASR, instead of always using enhanced speech for overlapping input.
In this paper, as an initial investigation for the switching between enhanced and observed signals, we propose a rule-based switching algorithm based on estimated SIR and SNR. 

We define the SIR and SNR as:
\begin{align}
    \label{eq:sir}
    \text{SIR} := 10 \log_{10} \frac{|| \mathbf{S} ||^{2}}{|| \mathbf{I} ||^{2}} ,\;\text{SNR} := 10 \log_{10} \frac{|| \mathbf{S} ||^{2}}{|| \mathbf{N} ||^{2}}
\end{align}
where $||\cdot||$ denotes the $L_{2}$ norm.
Based on SIR and SNR, the ASR input $\tilde{\mathbf{S}}$ is switched according to the following rule:
\begin{align}
    \label{eq:switch}
    \tilde{\mathbf{S}} = \begin{cases}
    \text{SE}(\mathbf{Y}, \mathbf{C}^{\text{S}}) & (f(\text{SIR}, \text{SNR}) < \lambda) \\
    \mathbf{Y} & (otherwise)
    \end{cases}
\end{align}
where $f(\cdot)$ denotes the conditional function and $\lambda$ denotes a threshold to control the input switching.
In this paper's experiment, we simply define the conditional function as follows:
\begin{align}
    \label{eq:score}
    f(\text{SIR}, \text{SNR}) := \text{SIR} - \text{SNR} = 10 \log_{10} \frac{|| \mathbf{N} ||^{2}}{|| \mathbf{I} ||^{2}}.
\end{align}
This rule reflects the observation that the enhanced speech tends to benefit ASR as the interfering speech $\mathbf{I}$ becomes more dominant, and that the observed mixture tends to benefit ASR as the noise $\mathbf{N}$ becomes more dominant.
Even if the interfering speaker is large to some extent, this rule suggests choosing observed mixtures as the input of ASR if noise is also large.
The way to estimate $f(\text{SIR}, \text{SNR})$ score is described in Section~\ref{section:sir_estimation}.

Although we investigated a simple rule-based switching to show the possibility of selective use of observed mixtures instead of enhanced speech, further improvement could be achieved by using more advanced rules or introducing learning-based switching (refer to Section \ref{section:evaluation} for more detailed explanations for prospects).

\subsection{Speech extraction-based SNR and SIR estimation\label{section:sir_estimation}}
In order to implement switching mechanism in Eq.~\eqref{eq:switch}, the score $f(\text{SIR}, \text{SNR})$ defined in Eq.~\eqref{eq:score} has to be estimated.

In this work, we assume that the speaker clue of the `interfering speaker' $\mathbf{C}^{\text{I}}$ is also available to estimate the interfering speech $\hat{\mathbf{I}}$ by applying target speech extraction expressed in Eq.~\eqref{eq:spkbeam}. 
For noise, we assume that the noise signal power is almost stable during a speech segment. 
We used the region of mixture where both target and interfering speakers are inactive for noise $L_{2}$ norm calculation.
We could eliminate the use of the interfering speaker clue by introducing noise prediction head to the SpeakerBeam structure, which is a part of our future works.

Since the duration of the region used for the calculation of $L_{2}$ norm are different between the interference and noise, we substituted $L_{2}$ norm in Eq.~\eqref{eq:score} with the average signal power over time $|| \cdot ||^{2}/T'$ where $T'$ denotes the length of the regions used for $L_{2}$ norm calculation. 
The active region of target and interference were estimated by multi-class voice activity detection explained in Section~\ref{section:enhancement-setup}.

\section{Experiments}
The experimental results will be presented in two parts: the analysis part and the evaluation of the switching part.
In the analysis part, we show the result of ASR performance comparison on observed mixture and enhanced speech at various SIR, SNR, and noise type combinations.
In the evaluation of the switching part, the ASR performance with the proposed switching method is presented and compared with those on observed mixture and enhanced speech.
The threshold for the switching in Eq.~\eqref{eq:switch} was experimentally determined as $\lambda = 10$ according to the character error rate (CER) for observed mixture and enhanced speech on the development set. 
For both parts, we evaluated CER as a speech recognition performance metric.
All experiments were performed on speech recordings simulated based on Corpus of Spontaneous Japanese (CSJ)\cite{maekawa2003corpus} and CHiME-4~\cite{chime4} corpora, sampled at 16 kHz.

\subsection{Target Speech Extraction\label{section:enhancement-setup}}
\textbf{Dataset}: For training of target speech extraction model and evaluation of target speech recognition pipeline, we created two-speaker and noise mixtures generated by mixing utterances and noise. 
For training, two-speaker mixtures are created by randomly selecting SIR values between -5 to 5 dB, in a similar way as widely used WSJ0-2mix corpus\cite{hershey2016deep}. 
To generate noisy mixtures, we randomly selected a noise signal out of four noise environments of CHiME-4 noise: street, cafe, bus, and pedestrian, and added it to every mixture at an SNR value randomly selected between 0 and 20 dB. 
For the target speaker clue, we randomly chose an utterance of the target speaker different from that in the mixture. 
The training set consisted of 50,000 mixtures from 937 speakers. The development set consisted of 5,000 mixtures from 49 speakers.
For evaluation, we created mixtures at various SIR between 0 and 20 dB and SNR between 0 and 20 dB for each type of noise environment.
Every evaluation dataset has 1,000 mixtures from 30 speakers, and each dataset has the same combination of speakers and utterances with only differences in SNR, SIR, and noise types.
The evaluation dataset consists of almost fully overlapping conditions with an overlap ratio of about 94 \% on average.

\textbf{Network configuration}: In this experiment, we adopted a time-domain speakerBeam structure\cite{luo2019conv, delcroix2020improving} as a speech extraction module in Eq.~\eqref{eq:spkbeam}. 
The time-domain SpeakerBeam conducts time-domain speech extraction, which has been shown to outperform time-frequency mask-based approaches. Thus, the input feature is a waveform.
In this work, we adopted multitask model of speech enhancement and multi-class voice activity detection (VAD) to further raise the performance of the speech extraction model\cite{zhuang2016multi}.
VAD head estimates the presence of target and interfering speaker respectively for each time frame from the output of the last dilated convolution block in the time-domain SpeakerBeam structure. 
The target was 4 classes of possible combinations of the presence of target and interfering speakers. 
Teacher labels for VAD were created by applying WebRTC VAD~\cite{webrtcvad} for original clean speeches before mixing.

\textbf{Training setup}: Training objective was scale-dependent source to distortion ratio (sd-SDR) loss for enhancement head and cross-entropy loss averaged over frames for VAD head. 
We set hyperparameters of the speakerBeam as follows: ${\rm N}=256$, ${\rm L}=20$, ${\rm B} = 256$,  ${\rm R}=3$, ${\rm X}=8$, ${\rm H}=512$ and ${\rm P} = 3$ following the notation in the original paper\cite{delcroix2020improving}. VAD head is composed of 2 layer BLSTM network with 128 cells followed by a linear layer.
For the optimization, we adopted the Adam algorithm\cite{kingma2014adam}. The initial learning rate was set as 5e-4 and we trained the models for 50 epochs and chose the best performing model based on a development set. Multitask weight was set as 1 for the speech enhancement loss and 10 for the VAD loss to make the range of each loss term roughly the same scale.

\subsection{ASR backend}
\textbf{Dataset}: In order to train the noise-robust speech recognition model, we created noisy CSJ dataset by adding CHiME-4 noise to the original CSJ corpus.
The data creation procedure of the clean speech is based on Kaldi CSJ recipe\cite{povey2011kaldi}.
For noise-robust training, we omitted cafe noise from CHiME-4 noise and randomly selected a noise signal out of the remaining three noise environments. 
This setting allows testing with unseen noise conditions. 
Since the adopted three noise environments do not contain much background speech compared with cafe noise, the ASR model trained here is not particularly robust to interference speakers, and thus speech extraction would play an important role for this setup.
The SNR was randomly selected between 0 and 20 dB.
Besides, we convolved randomly generated simulated room impulse responses (RIRs) generated by image method \cite{allen1979image} for 90\% of the training and validation dataset to make the ASR model closer to those used in practical situations.
T60 was randomly selected between 150 and 600 ms and the distance between a microphone and a speaker was randomly selected between 10 and 300 cm.

\textbf{Network configuration and training setup}: We used ESPnet to build the ASR backend model\cite{watanabe2018espnet}, which is an open-source toolkit for end-to-end speech processing.
We adopted transformer-based encoder-decoder ASR model with connectionist temporal classification objective\cite{karita2020ctctransformer}.
The training was conducted according to the default setup of ESPnet CSJ recipe\cite{espnetrecipe} with speed perturbation\cite{povey2015sp} and SpecAugment\cite{specaugment}, except that we did not use language model shallow fusion.

\begin{table*}[tbh]
\centering
\caption{Relative CER reduction of target speech extraction from the observed mixture. Positive values indicate that the speech extraction improved ASR performance and negative values indicate that speech extraction degreaded ASR performance. Under severe noise condition like SNR 0 dB, speech extraction sometimes did not improve ASR performance even if the interfering speech present at a moderate level of SIR 10 dB. }
\label{tab:analysis}
\begin{tabular}{r|rrrrrrrrrrrr}
\hline
\multicolumn{1}{c|}{\multirow{3}{*}{\begin{tabular}[c]{@{}c@{}}SIR\\ {[}dB{]}\end{tabular}}} & \multicolumn{12}{c}{Noise Type and   SNR {[}dB{]}} \\
\multicolumn{1}{c|}{} & \multicolumn{3}{c|}{CAF} & \multicolumn{3}{c|}{PED} & \multicolumn{3}{c|}{STR} & \multicolumn{3}{c}{BUS} \\
\multicolumn{1}{c|}{} & \multicolumn{1}{c}{20} & \multicolumn{1}{c}{10} & \multicolumn{1}{c|}{0} & \multicolumn{1}{c}{20} & \multicolumn{1}{c}{10} & \multicolumn{1}{c|}{0} & \multicolumn{1}{c}{20} & \multicolumn{1}{c}{10} & \multicolumn{1}{c|}{0} & \multicolumn{1}{c}{20} & \multicolumn{1}{c}{10} & \multicolumn{1}{c}{0} \\ \hline
0 & 89\% & 86\% & \multicolumn{1}{r|}{57\%} & 90\% & 85\% & \multicolumn{1}{r|}{49\%} & 90\% & 87\% & \multicolumn{1}{r|}{67\%} & 90\% & 89\% & 80\% \\
5 & 83\% & 74\% & \multicolumn{1}{r|}{20\%} & 83\% & 72\% & \multicolumn{1}{r|}{7\%} & 84\% & 77\% & \multicolumn{1}{r|}{36\%} & 84\% & 81\% & 62\% \\
10 & 64\% & 37\% & \multicolumn{1}{r|}{{\ul -25\%}} & 62\% & 31\% & \multicolumn{1}{r|}{{\ul -41\%}} & 66\% & 46\% & \multicolumn{1}{r|}{{\ul -22\%}} & 68\% & 57\% & 18\% \\
15 & 28\% & {\ul -11\%} & \multicolumn{1}{r|}{{\ul -47\%}} & 27\% & {\ul -21\%} & \multicolumn{1}{r|}{{\ul -53\%}} & 33\% & 7\% & \multicolumn{1}{r|}{{\ul -61\%}} & 39\% & 20\% & {\ul -23\%} \\
20 & {\ul -27\%} & {\ul -52\%} & \multicolumn{1}{r|}{{\ul -50\%}} & {\ul -29\%} & {\ul -68\%} & \multicolumn{1}{r|}{{\ul -57\%}} & {\ul -15\%} & {\ul -41\%} & \multicolumn{1}{r|}{{\ul -75\%}} & {\ul -6\%} & {\ul -23\%} & {\ul -43\%} \\ \hline
\end{tabular}
\end{table*}

\begin{table}[tbh]
\centering
\caption{CER for (a) enhanced speech, (b) observed mixture and (c) proposed switching method between (a) enhanced and (b) mixture. (d) shows the relative improvement of (c) proposed switching from (a) enhanced speech. The switching rule is supposed to choose observed mixture for shaded conditions in (c) and (d).}
\label{tab:result}
\scalebox{0.95}[0.95]{
\begin{tabular}{rrrrlrrrr}
\multicolumn{4}{l}{(a) enhanced speech} &  & \multicolumn{4}{l}{(b) observed mixture} \\ \cline{1-4} \cline{6-9} 
\multicolumn{1}{c|}{} & \multicolumn{3}{c}{SNR {[}dB{]}} &  & \multicolumn{1}{c|}{} & \multicolumn{3}{c}{SNR {[}dB{]}} \\
\multicolumn{1}{c|}{\multirow{-2}{*}{\begin{tabular}[c]{@{}c@{}}SIR\\ {[}dB{]}\end{tabular}}} & \multicolumn{1}{c}{20} & \multicolumn{1}{c}{10} & \multicolumn{1}{c}{0} &  & \multicolumn{1}{c|}{\multirow{-2}{*}{\begin{tabular}[c]{@{}c@{}}SIR\\ {[}dB{]}\end{tabular}}} & \multicolumn{1}{c}{20} & \multicolumn{1}{c}{10} & \multicolumn{1}{c}{0} \\ \cline{1-4} \cline{6-9} 
\multicolumn{1}{r|}{0} & \textbf{8.4} & \textbf{10.9} & \textbf{31.2} &  & \multicolumn{1}{r|}{0} & 80.7 & 80.6 & 83.9 \\
\multicolumn{1}{r|}{5} & \textbf{6.7} & \textbf{9.0} & \textbf{28.2} &  & \multicolumn{1}{r|}{5} & 40.9 & 37.5 & 40.1 \\
\multicolumn{1}{r|}{10} & \textbf{5.8} & \textbf{7.8} & 24.9 &  & \multicolumn{1}{r|}{10} & 16.5 & 13.8 & \textbf{20.4} \\
\multicolumn{1}{r|}{15} & \textbf{5.4} & 7.6 & 23.6 &  & \multicolumn{1}{r|}{15} & 7.9 & \textbf{7.5} & \textbf{15.9} \\
\multicolumn{1}{r|}{20} & 6.3 & 8.9 & 23.2 &  & \multicolumn{1}{r|}{20} & \textbf{5.3} & \textbf{6.0} & \textbf{14.8} \\ \cline{1-4} \cline{6-9} 
\multicolumn{1}{l}{} & \multicolumn{1}{l}{} & \multicolumn{1}{l}{} & \multicolumn{1}{l}{} &  & \multicolumn{1}{l}{} & \multicolumn{1}{l}{} & \multicolumn{1}{l}{} & \multicolumn{1}{l}{} \\
\multicolumn{4}{l}{\begin{tabular}[c]{@{}l@{}}(c) proposed\\ switching between\\ enhanced and mixture\end{tabular}} &  & \multicolumn{4}{l}{\begin{tabular}[c]{@{}l@{}}(d) relative improvement\\  of (c) switching \\  from (a) enhanced speech\end{tabular}} \\ \cline{1-4} \cline{6-9} 
\multicolumn{1}{c|}{} & \multicolumn{3}{c}{SNR {[}dB{]}} &  & \multicolumn{1}{c|}{} & \multicolumn{3}{c}{SNR {[}dB{]}} \\
\multicolumn{1}{c|}{\multirow{-2}{*}{\begin{tabular}[c]{@{}c@{}}SIR\\ {[}dB{]}\end{tabular}}} & \multicolumn{1}{c}{20} & \multicolumn{1}{c}{10} & \multicolumn{1}{c}{0} &  & \multicolumn{1}{c|}{\multirow{-2}{*}{\begin{tabular}[c]{@{}c@{}}SIR\\ {[}dB{]}\end{tabular}}} & \multicolumn{1}{c}{20} & \multicolumn{1}{c}{10} & \multicolumn{1}{c}{0} \\ \cline{1-4} \cline{6-9} 
\multicolumn{1}{r|}{0} & 8.4 & 10.9 & 31.2 &  & \multicolumn{1}{r|}{0} & 0\% & 0\% & 0\% \\
\multicolumn{1}{r|}{5} & 6.7 & 9.0 & 28.4 &  & \multicolumn{1}{r|}{5} & 0\% & 0\% & {\ul -1\%} \\
\multicolumn{1}{r|}{10} & 5.8 & 7.8 & \cellcolor[HTML]{D9D9D9}21.1 &  & \multicolumn{1}{r|}{10} & 0\% & 0\% & \cellcolor[HTML]{D9D9D9}\textbf{15\%} \\
\multicolumn{1}{r|}{15} & 5.4 & 7.3 & \cellcolor[HTML]{D9D9D9}17.3 &  & \multicolumn{1}{r|}{15} & 0\% & \textbf{4\%} & \cellcolor[HTML]{D9D9D9}\textbf{27\%} \\
\multicolumn{1}{r|}{20} & 6.1 & \cellcolor[HTML]{D9D9D9}7.3 & \cellcolor[HTML]{D9D9D9}17.0 &  & \multicolumn{1}{r|}{20} & \textbf{2\%} & \cellcolor[HTML]{D9D9D9}\textbf{18\%} & \cellcolor[HTML]{D9D9D9}\textbf{27\%} \\ \cline{1-4} \cline{6-9} 
\end{tabular}
}
\end{table}

\subsection{Experimental results}
\subsubsection{Analysis: Observed mixture vs Enhanced speech for different SIR and SNR ratios \label{section:analysis}}
Table~\ref{tab:analysis} shows the relative CER reduction of speech extraction from the mixture. 
For low SIR conditions as 0 dB, the speech enhancement technique benefits ASR performance drastically by reducing CER 80\% compared with the observed mixture on average.
On the other hand, for high SIR conditions as 20 dB, speech enhancement degraded speech recognition performance. This result supports the empirical knowledge that ASR has an ability to recognize dominant speech under the presence of interfering speech.
In addition, as non-speech noise becomes more dominant, the enhanced speech becomes more detrimental for ASR.
It should be noted that under severe noise condition as SNR 0 dB, speech extraction did not improve ASR performance even if the interfering speech is present at a moderate level of SIR 10 dB in some conditions. 

The result also shows that whether the enhancement technique benefits ASR or not also depends on noise types.
For example, the enhancement benefit ASR for street and bus noise but degrade ASR for cafe and pedestrian noise, when SIR is 15 dB and SNR is 10 dB.

The results shown here indicate that we should not apply speech extraction for all the overlapping speech to get better ASR results. 

\subsubsection{Evaluation: Estimation-based input signal switching \label{section:evaluation}}
We evaluated out proposed switching method of the ASR input between the observed mixture and enhanced speech.
Table~\ref{tab:result} shows the CER of (a) enhanced speech, (b) observed mixture, (c) the proposed switching method, and (d) the relative improvement of CER derived by the switching method from enhanced speech. The bold numbers in (a) and (b) indicate which one of enhanced speech or observed mixture is best for ASR.
The results shown here are average values over noise types.
Based on the threshold $\lambda = 10$, the adopted switching rule is supposed to choose `observed mixture' for ASR input when $\text{SIR} - \text{SNR} \geq 10\,\text{dB}$, which corresponds to shaded conditions in (c) and (d). 
In (d), results displayed in boldface are conditions where the switching brought improvement, and underlined results are conditions where the switching degraded ASR performance.

According to the results shown in (c), the proposed switching mechanism significantly reduced CER for $\text{SIR} - \text{SNR} \geq 10\,\text{dB}$ conditions from that for (a) enhanced speech. 
The average CER reduction for $\text{SIR} - \text{SNR} \geq 10\,\text{dB}$ conditions were 22 \%. 
Although the switching rule is supposed to choose enhanced speech for conditions like SIR 20 dB and SNR 20 dB based on the rule, the estimation errors of SIR and SNR make the switching method select observed mixtures, which unintentionally improved the ASR performance. (Note that the switching rule is defined based on the experimental observations for the development set and not necessarily optimal for the evaluation set.)
At the same time of gaining improvement, the ASR result degradation were also very limited for $\text{SIR} - \text{SNR} < 10\,\text{dB}$ conditions.
These results suggested that the switching mechanism worked well even with the estimated SIR and SNR scores.

In case the switching worked perfectly, the achieved CER would be the minimum value of (a) enhanced speech and (b) observed mixture.
Since the achieved CER still has a gap from such oracle values, it seems that there remains room for further improvement of speech recognition performance for overlapping speech.
As prospects, we can improve SIR and SNR prediction methods as well as the switching rules into more advanced ones, including learning-based SNR and SIR prediction, rules considering noise types, frame-level switching, etc.
It is also possible to adopt joint training frameworks to train neural network-based switching modules on ASR loss, in order to make the switching module acquire selection criteria that maximize ASR performance.

\section{Conclusion}
In this paper, we investigated in which conditions speech separation/extraction were beneficial for ASR. 
We observed that depending on the SIR and SNR conditions, better ASR performance could be obtained by recognizing the observed mixture signal instead of the enhanced speech.
As an initial investigation for the switching between enhanced and observed signals, we proposed a rule-based switching mechanism based on estimated SIR and SNR. 
It is experimentally confirmed that even a simple switching rule could already improve ASR performance by up to 27 \% relative CER reduction. 
We expect that more advanced switching mechanisms could further improve the performance.

\pagebreak

\bibliographystyle{IEEEtran}

\tiny{
\bibliography{mybib}
}
\end{document}